\documentstyle[11pt,newpasp,twoside,epsf]{article}
\begin{document}

\title{Constraints on Cluster Formation from Old Globular Cluster Systems}

\author{Dean E. McLaughlin\altaffilmark{1}}

\affil{Department of Astronomy, 601 Campbell Hall, University of California,
Berkeley, CA 94720}

\altaffiltext{1}{Hubble Fellow}

\begin{abstract}
The properties of old globular cluster systems (GCSs) in galaxy halos offer
unique insight into the physical processes that conspire to form any generic
star cluster, at any epoch. Presented here is a summary of the information
obtained from (1) the specific frequencies (total populations) and spatial
structures (density vs.~galactocentric radius) of GCSs in early-type galaxies,
as they relate to the efficiency (or probability) of bound cluster
formation, and (2) the fundamental role of a scaling between cluster mass and
energy among Galactic globulars in setting their other structural correlations,
and the possible implications for star formation efficiency as a function of
mass in gaseous protoclusters.
\end{abstract}

\keywords{galaxies: star clusters -- globular clusters: general --
stars: formation}

\section{Introduction}

Until quite recently, it was commonly assumed that the old globular clusters
in galaxy halos were the remnants of a unique sort of star formation that
occurred only in a cosmological context. The discovery of young, massive,
``super'' star clusters in local galaxy mergers and starbursts has clearly
done much to change this perception; but at least as important is the parallel
recognition that star formation in the Milky Way itself proceeds---under
much less extreme conditions---largely in a clustered mode. Observations of
entire starbursts (Meurer et al.~1995) and individual Galactic molecular
clouds (e.g., Lada 1992), as well as a more general comparison of the mass
function of molecular cloud clumps and the stellar IMF (Patel \& Pudritz 1994),
all argue convincingly that (by mass) most new stars are born in groups rather
than in isolation. The production of a true stellar cluster---one that
remains bound even after dispersing the gas from which it formed---is
undoubtedly a {\it rare} event, but it is an exceedingly {\it regular} one.

Seen in this light, the globular cluster systems (GCSs) found in most galaxies
can be used to good effect as probes not only of galaxy formation but also
of an important element of the generic star-formation process at any epoch.
This is arguably so even in cases where newly formed clusters may not be
``massive'' according to the criteria of this workshop (the main issue
being simply the formation of a {\it self-gravitating} stellar system), and
even though GCSs have been subjected to $10^{10}$ yr of dynamical evolution in
the tidal fields of their parent galaxies (see O.~Gerhard's contribution to
these proceedings, and note that theoretical calculations geared specifically
to conditions both in the Milky Way [Gnedin \& Ostriker 1997] and in the giant
elliptical M87 [Murali \& Weinberg 1997] suggest that GCS properties are most
affected by evolution inside roughly a stellar effective radius in each case).

\section{The Efficiency of Cluster Formation}

At some point during the collapse and fragmentation of a cluster-sized
cloud of gas, the massive stars which it has formed will expel any remaining
gas by the combined action of their stellar winds, photoionization, and
supernova explosions. If the star formation efficiency
of the cloud, $\eta\equiv M_{\rm stars}/(M_{\rm stars}+M_{\rm gas})$, is below
a critical threshold just when the gas is lost, then the blow-out removes
sufficient energy that the stellar group left behind is unbound and disperses
into the field. The precise value of this threshold depends on details of the
internal density and velocity structure of the initial gas cloud, and
on the timescale over which the massive stars dispel the gas; but various
estimates place it in the range $\eta_{\rm crit}\sim 0.2$--0.5 (e.g., Hills
1980; Verschueren 1990; Goodwin 1997, and these proceedings). There is no
theory which can predict whether any given piece of gas can ultimately achieve
$\eta>\eta_{\rm crit}$, but it is straightforward to evaluate {\it
empirically} the frequency---or efficiency---with which this occurs.

Traditionally, this has been discussed for GCSs in terms of {\it specific
frequency}, defined by Harris \& van den Bergh (1981) as the normalized ratio
of a galaxy's total GCS population to its $V$-band luminosity:
$S_N \equiv {\cal N}_{\rm tot}\times 10^{0.4(M_V+15)}$. As is well known
(see, e.g., Elmegreen 2000 for a recent review), there are
substantial and systematic variations in this ratio from one galaxy to
another: Global specific frequencies {\it decrease} with increasing galaxy
luminosity for early-type dwarfs, then {\it increase} gradually with
$L_{V,{\rm gal}}$ in normal giant ellipticals, and finally increase rapidly
with galaxy luminosity among the central ellipticals (BCGs) in groups and
clusters of galaxies. In addition, the more extended spatial distribution
of GCSs relative to halo stars in some (but not all) bright ellipticals leads
to {\it local} specific frequencies (ratios of GCS and field-star densities)
that increase with radius inside the galaxies (see McLaughlin 1999).

However, McLaughlin (1999) shows (following related work by Blakeslee et
al.~1997 and Harris et al.~1998) that these trends in
$S_N$ do {\it not} reflect any such behavior in the ability
to form globulars in protogalaxies. To see this, it is best to work in terms
of an efficiency per unit {\it mass}, $\epsilon_{\rm cl}\equiv
M_{\rm gcs}^{\rm init}/M_{\rm gas}^{\rm init}$,
where $M_{\rm gas}^{\rm init}$ is the total gas supply that was available to
form stars in a protogalaxy (whether in a monolithic collapse or a slower
assembly of many distinct, subgalactic clumps is unimportant) and
$M_{\rm gcs}^{\rm init}$ is the total mass of all globulars formed in that
gas. As McLaughlin (1999) argues, the integrated mass of an entire GCS should
{\it not} be much affected by dynamical evolution, and it is most appropriate
to include any gas presently associated with galaxies, as well as their
stellar masses, in estimating their initial gas contents. The {\it observable}
ratio $M_{\rm gcs}/(M_{\rm gas}+M_{\rm stars})$ should therefore
improve on $S_N\propto M_{\rm gcs}/M_{\rm stars}$ as an estimator of
$\epsilon_{\rm cl}$.

\begin{figure}[!b]
\plotfiddle{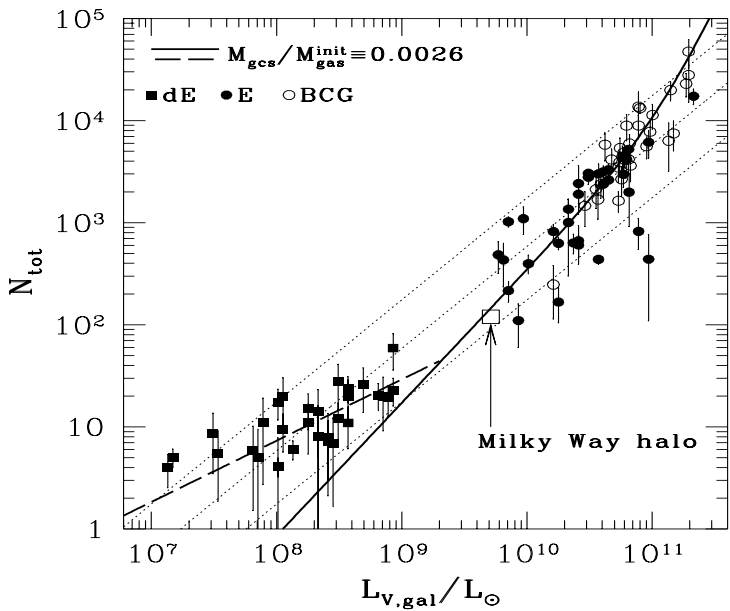}{3.0truein}{0}{50}{40}{-150}{-60}
\caption{\rm Constant efficiency of cluster formation,
$\epsilon_{\rm cl}\equiv 0.0026$ (bold lines) in 97 early-type systems and the
spheroid of the Milky Way. Light, dotted lines represent constant specific
frequencies ($S_N=8.55\times 10^7\,{\cal N}_{\rm tot}/L_{V,{\rm gal}}$) of
15, 5, and 1.5. From McLaughlin (1999).
\label{fig1}}
\end{figure}

Figure \ref{fig1} shows the total GCS populations vs.~galaxy luminosity in
97 early-type galaxies and the metal-poor spheroid of the Milky Way and
compares the expectations for a {\it constant} $\epsilon_{\rm cl}=
0.26\%$, given both the variation of stellar mass-to-light ratio with
$L_{V,{\rm gal}}$ on the fundamental plane of ellipticals and the
increase of $M_{\rm gas}/M_{\rm stars}$ with $L_{V,{\rm gal}}$ for regular
gE's and BCGs inferred from the correlation between their X-ray and optical
luminosities (bold solid curve; see McLaughlin 1999), and after correcting
(according to the model of Dekel \& Silk 1986) for the gas mass lost in
supernova-driven winds from early bursts of star formation in faint dwarfs
($L_{V,{\rm gal}}\le 2\times 10^9\,L_\odot$; bold dashed line). All {\it
systematic} variations in GCS specific frequencies reflect only different
relations, in different magnitude ranges, between $M_{\rm gas}^{\rm init}$ and
the present-day $L_{V,{\rm gal}}$.

McLaughlin (1999) also shows that the ratio of local densities,
$\rho_{\rm gcs}/(\rho_{\rm gas}+\rho_{\rm stars})$, is constant as a function
of galactocentric position (beyond a stellar effective radius) in each of the
large ellipticals M87, M49, and NGC 1399, and that this ratio is the same
in all three systems: $\epsilon_{\rm cl}=0.0026\pm0.0005$. Moreover, it seems
(although the data are much less clear in this case) that the same efficiency
also applies to the ongoing formation of open clusters in the Galactic disk.
It therefore appears that there is a {\it universal efficiency for cluster
formation}, whose value should serve as a strong constraint on very general
theories of star formation. (Note that one exception to the figure of
0.26\% by mass {\it may} be the formation of massive clusters in
mergers and starbursts, where it has been suggested that $\epsilon_{\rm cl}
\sim1$--10\% [e.g., Meurer et al.~1995; Zepf et al.~1999]. However, this
conclusion is very uncertain and requires more careful investigation.)

While this result certainly has interesting implications for aspects of
large-scale galaxy formation (McLaughlin 1999; Harris et al.~1998),
the main point to be emphasized here is that the variations in early-type
GCS specific frequencies are now understood to result from
variations in the gas-to-star mass ratio in galaxies, rather than from any
peculiarities in their GCS abundances per se (cf.~the similar suggestion of
Blakeslee et al.~1997). That is, the efficiency of {\it unclustered} star
formation was {\it not} universal in protogalaxies: while globulars apparently
always formed in just the numbers expected of them, the formation of a normal
proportion of field stars was subsequently disabled in many cases. The clumps
of gas which formed bound clusters therefore must have
collapsed before those forming unbound groups and associations, i.e.,
they must have been denser than average. This and the insensitivity of
$\epsilon_{\rm cl}$ to local or global galaxy environment together suggest that
quantitative theories of cluster formation should seek to identify a {\it
threshold in relative density}, $\delta\rho/\rho$, that is always exceeded
by $\simeq$0.26\% of the mass fluctuations in any large star-forming complex.

\section{Globular Cluster Binding Energies}

Even as they clarify the {\it probability} that a $\sim$$10^5$--$10^6\,M_\odot$
clump of gas was able to form stars with cumulative efficiency $\eta$ high
enough to produce a bound globular cluster, the integrated GCS mass ratios in
galaxies say nothing of {\it how} this was achieved in any individual case.
This more ambitious question is essentially one of energetics---When does the
energy injected by the massive stars in an embedded young cluster overcome the
binding energy of whatever gas remains, thus expelling it and terminating star
formation?---and its answer requires both an understanding of local
star formation laws ($d{\rho}_{\rm stars}/dt$ as a function of
$\rho_{\rm gas}$) and a self-consistent treatment of feedback on small
($\sim$10--100 pc) scales. One way to begin addressing this complex problem
empirically is to compare the energies of globular clusters with the initial
energies of their gaseous progenitors.

McLaughlin (2000) has calculated the $V$-band mass-to-light ratios of 39
regular (non--core-collapsed) Milky Way globulars, and finds that they are all
consistent with a single $\Upsilon_{V,0}=(1.45\pm0.10)\ M_\odot\,L_\odot^{-1}$.
Applying this to all other Galactic globulars, and adopting single-mass,
isotropic King (1966) models for their internal structure, then allows binding
energies $E_b$ to be estimated for a complete sample of 109 regular (and
30 post--core-collapse) objects. This exercise reveals a very tight
correlation between $E_b$, total cluster luminosity $L$ (or mass
$M=\Upsilon_{V,0}L$), and Galactocentric position: $E_b=7.2\times10^{39}\
{\rm erg}\,(L/L_\odot)^{2.05}\,(r_{\rm gc}/8\,{\rm kpc})^{-0.4}$, with
uncertainties of roughly $\pm$0.1 in each of the fitted exponents on $L$
and $r_{\rm gc}$ (cf.~Saito 1979, who claimed $E_b\propto M^{1.5}$ on the
basis of a much smaller dataset).

\begin{figure}[!b]
\plotfiddle{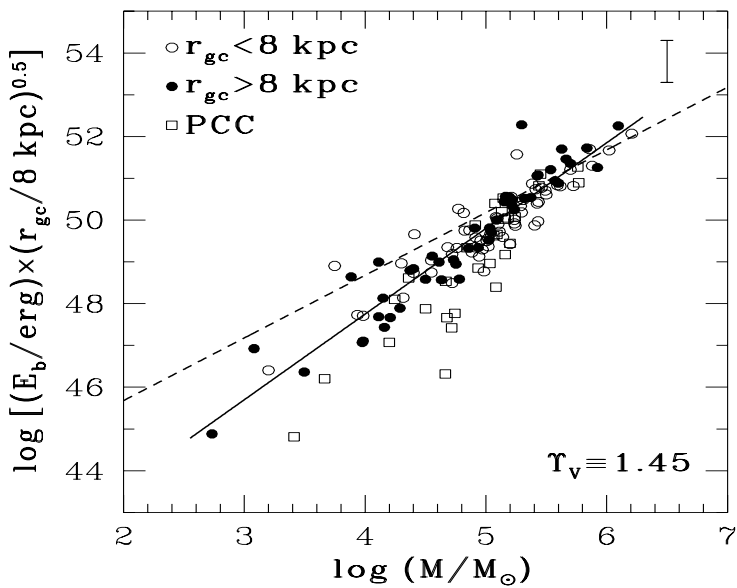}{3.0truein}{0}{50}{40}{-150}{-60}
\caption{Binding energy vs.~mass for globular clusters (points and solid
line; see McLaughlin 2000) and their gaseous progenitors (broken line) in the
Galaxy. Total cluster luminosities are converted to masses by applying the
constant mass-to-light ratio indicated.
\label{fig2}}
\end{figure}

These constraints on $\Upsilon_{V,0}$ and for $E_b(L,r_{\rm gc})$ are, in
fact, two edge-on views of a {\it fundamental plane} in the (four-dimensional)
parameter space of King models, to which real globulars are confined in the
Milky Way (cf.~Djorgovski 1995; Bellazzini 1998). The full characteristics of
this plane subsume {\it all other observable correlations} between any
combination of other cluster parameters (see McLaughlin 2000), and they
therefore provide a {\it complete} set of independent facts to be explained in
any theory of globular cluster formation and evolution. In fact, the
$E_b$--$L$ correlation is stronger among clusters at larger Galactocentric
radii (where dynamical cluster evolution is weaker), suggesting that it was
set largely by the cluster {\it formation} process. The same is true of a
weaker correlation between cluster concentration and luminosity (see Vesperini
1997), which is related to the distribution of globulars {\it on} the
fundamental plane.

Any collection of critically stable, virialized gas spheres under a surface
pressure $P_s$ have a common column density, $\Sigma\equiv M/(\pi R^2)
\propto P_s^{0.5}$, and thus $E_b^{\rm gas}\equiv GM^2/R \propto M^{1.5}
P_s^{0.25}$. Harris \& Pudritz (1994) have developed a physical framework in
which protoglobular clusters in the Milky Way were massive analogues of the
dense clumps in disk molecular clouds today; in particular, their column
densities were the same: $\Sigma\simeq 10^3\ M_\odot$ pc$^{-2}$ at
$r_{\rm gc}=8$ kpc. In addition, it is natural to expect $P_s \propto
r_{\rm gc}^{-2}$ for such protocluster clumps embedded in larger (but
subgalactic) star-forming clouds that were themselves surrounded by a diffuse
medium virialized in a ``background'' isothermal potential well (Harris \&
Pudritz 1994). Together, these basic hypotheses imply $E_b^{\rm gas}=4.8\times
10^{42}\ {\rm erg}\,(M/M_\odot)^{1.5}\,(r_{\rm gc}/8\,{\rm kpc})^{-0.5}$.
Note that the $r_{\rm gc}$ scaling is essentially that observed directly for
Galactic globulars today, enabling a direct comparison of the (model)
initial and final $E_b(M,r_{\rm gc})$ relations in Fig.~\ref{fig2}.

To explain the relative $E_b(M)$ normalizations in Fig.~\ref{fig2}
requires quantitative modelling of the initial structure and feedback dynamics
in the gaseous protoclusters. Meanwhile, the different {\it slopes} of the
two relations are significant: The ratio of the initial energy of a gaseous
clump to the final $E_b$ of a stellar cluster is a non-decreasing
function of the cumulative star formation efficiency $\eta$; but this Figure
shows that it is also an increasing function of cluster mass, and thus that
$\eta$ {\it was systematically higher in more massive protoclusters}. The 
quantitative details of this dependence are also model-dependent (McLaughlin,
in preparation), but the inference on the qualititative behavior of $\eta$ is
robust and presents a new constraint for theories of cluster formation.
Once the behavior of $\eta$ as a function of initial gas mass is understood,
progress will have been made in explaining the universal $\epsilon_{\rm cl}$ of
\S2, and there will be further implications for other global properties of
GCSs---such as their mass functions, which, contrary to current modelling
(McLaughlin \& Pudritz 1996; Elmegreen \& Efremov 1997), can no longer simply
be assumed proportional to those of their gaseous protoclusters.

\begin{acknowledgements}

This work was supported by NASA through grant number HF-1097.01-97A awarded by
the Space Telescope Science Institute, which is operated by the Association of
Universities for Research in Astronomy, Inc., for NASA under contract
NAS5-26555.

\end{acknowledgements}

\appendix\section*{Discussion}

\noindent{\bf G.~Meurer:}
Concerning the two-orders-of-magnitude difference between
$\epsilon_{\rm cl}$ and the fraction of UV light in starbursts: One order of
magnitude may be explainable by the gas content in starbursts.

\

\noindent{\bf McLaughlin:}
That does seem plausible (e.g., Zepf et al.~1999), although it
should of course be checked in detail in every individual case. But the gas
mass in starbursts really does have to enter as much more than a factor-of-ten
effect if there is no boost in the cluster formation efficiency
in starbursts vs.~old galaxy halos. A real question remains as to whether or
not that is the case.

\

\noindent{\bf G.~\"Ostlin:}
Since none of the fundamental properties of globular clusters depend on
metallicity, including the core mass-to-light ratio which appears constant,
I guess this requires them to have had a universal stellar IMF, independent
of metallicity.

\

\noindent{\bf McLaughlin:}
I think that's exactly right.

\end{document}